\documentclass[twocolumn]{IEEEtran}

\usepackage{amsmath, graphics,amssymb,epsfig,subfigure}
\usepackage{algorithm}
\usepackage{algorithmic}
\usepackage{float}
\usepackage{multirow}
\usepackage{cite}

\newlength{\capwidth}
\newtheorem{Theorem}{Theorem}
\newtheorem{Corollary}{Corollary}

\newtheorem{Lemma}{Lemma}

\newcommand{\fv}{\mbox{$\bf f $}}
\newcommand{\Yv}{\mbox{$\bf Y $}}
\newcommand{\Hv}{\mbox{$\bf H $}}

\newcommand{\Av}{\mbox{$\bf A $}}
\newcommand{\xv}{\mbox{$\bf x $}}

\newcommand{\nv}{\mbox{$\bf n $}}

\newcommand{\hv}{\mbox{$\bf h $}}
\newcommand{\ev}{\mbox{$\bf e $}}

\newcommand{\Wv}{\mbox{$\bf W $}}

\newcommand{\Iv}{\mbox{$\bf I $}}

\newcommand{\Kv}{\mbox{$\bf K $}}

\newcommand{\Fv}{\mbox{$\bf F $}}

\newcommand{\Lv}{\mbox{$\bf L $}}
\newcommand{\Gv}{\mbox{$\bf G $}}

\newcommand{\Zv}{\mbox{$\bf Z $}}
\newcommand{\vv}{\mbox{$\bf v $}}

\newcommand{\be}{\begin{equation}}
\newcommand{\eq}{\end{equation}}
\newcommand{\ee}{\end{equation}}
\newcommand{\bea}{\begin{eqnarray}}
\newcommand{\eea}{\end{eqnarray}}
\newcommand{\bdp}{\begin{displaymath}}
\newcommand{\edp}{\end{displaymath}}
\begin{document}
\title{\huge{MIMO Broadcasting for Simultaneous Wireless Information and Power Transfer: Weighted MMSE Approaches}}

\author{\IEEEauthorblockN{Changick Song, Cong Ling, Jaehyun Park, and Bruno Clerckx} \\
\thanks{C. Song is with the Dept. of Information and Communications Eng., Korea National University of Transportation, Chungju, Korea
(e-mail: c.song@ut.ac.kr).

J. Park is with the Dept. of Electronics Eng., Pukyong National University, Busan, Korea (e-mail: jaehyun@pknu.ac.kr).

C. Ling is with the Dept. of Electrical and Electronic Eng., Imperial College, London, UK (e-mail: c.ling@imperial.ac.uk).

B. Clerckx is with the Dept. of Electrical and Electronic Eng., Imperial College, London, UK and the School of Electrical Eng., Korea University, Seoul, Korea
(e-mail: b.clerckx@imperial.ac.uk).

}
}\maketitle

\begin{abstract}
We consider simultaneous wireless information and power transfer (SWIPT) in MIMO Broadcast networks
where one energy harvesting (EH) user and one information decoding (ID) user share the same time and frequency resource.
In contrast to previous SWIPT systems based on the information rate, this paper addresses the problem
in terms of the weighted minimum mean squared error (WMMSE) criterion.
First, we formulate the WMMSE-SWIPT problem which minimizes the weighted sum-MSE of the message signal arrived at the ID user,
while satisfying the requirement on the energy that can be harvested from the signal at the EH user.
Then, we propose the optimal precoder structure of the problem and identify the best possible MSE-energy tradeoff region
through the alternative update of the linear precoder at the transmitter with the linear receiver at the ID user.
From the derived solution, several interesting observations are made compared to the conventional SWIPT designs.
\end{abstract}

\vspace{-0pt}
\section{Introduction}\label{sec:introduction}

In recent years, it has been recognized that radio frequency (RF) signals
that transport information can at the same time be exploited as significant energy
source for devices that are capable of harvesting RF energy to power themselves.
For this reason, {\it simultaneous wireless information and power transfer} (SWIPT)
has appeared as promising technologies in conjunction with energy harvesting (EH) devices
for providing energy constrained wireless networks with convenient and perpetual energy supplies \cite{GPulkit:10} \cite{LVarshney:08}.

In SWIPT systems, efficient transmitter designs have been active research area over the last few years
as in \cite{JPark:13,ANasir:12,JXu:13,HSon:14,RZhang:13} and references therein.
For example, various beamforming strategies have been proposed in multi-antenna networks such as interference
\cite{JPark:13}, relaying \cite{ANasir:12}, and broadcast channels (BC) \cite{JXu:13} \cite{HSon:14}
to maximize the rate at the information decoding (ID) users and the harvested energy at the EH users simultaneously.
Especially in \cite{RZhang:13}, the authors completely characterized the rate-energy tradeoff region by suggesting the optimum precoder for
multiple-input multiple-output (MIMO) two-user BC SWIPT where one EH user and one ID user share the same spectrum resource.

While such information rate based designs may be useful for evaluating the system performance,
it may not be realistically implementable due to high encoding and decoding complexity.
As a low complexity alternative,
linear receivers such as the minimum mean squared error (MMSE) have also been widely proposed
in conventional communication networks where the information transfer was the only concern \cite{Sampath:01,Palomar:03,CSong:09TWC}.
Surprisingly, however, there is no earlier contribution on the linear receiver issue in the context of SWIPT systems.
As an initial step, our work develops a new design method of the transmitter and ID receiver in terms of the MMSE
for MIMO BC SWIPT where an EH user and an ID user are located separately, and
identify the best possible MSE-energy tradeoff region.

To get a better handle on the stream-wise error performance, we formulate a weighted MMSE (WMMSE) SWIPT problem which minimizes the weighted sum errors for the ID user
while satisfying the predetermined EH constraint for the EH user.
The weight matrix makes the problem generally non-convex; thus more challenging than the equal weight cases.
To solve the problem, we first focus on the transmitter side
for a given ID receiver and derive the solution as a semi-closed-form that can be obtained by simple bisection methods.
It is then observed that when the ID user has a single antenna, the derived precoder reduces to the optimal beamformer that
maximizes the signal-to-noise ratio (SNR) corresponding to the maximum information rate scheme in \cite{RZhang:13}.
It is particularly interesting to observe that while the proposed solution is attainable by a simple bisection method,
the previous design in \cite{RZhang:13} still needs to compute complicated ellipsoid methods to arrive at the same point.

Meanwhile, for a given the precoder, the optimal ID receiver is simply the Wiener filter \cite{Joham:05}.
Thus, we can easily optimize the transceiver through the alternative update process.
Due to the non-convexity of the problem, our solution may not ensure the global optimality, but
multiple initial points may take us closer to the optimal point.
Thus, the resulting MSE-energy region will carry a significant meaning as an MMSE counterpart of the rate-energy region in \cite{RZhang:13}.
Simulation results show that our solution exhibits improved bit error rate (BER) performance and convergence rate
compared to the conventional rate-based designs.
In addition, when we set the weight matrix to be equal to the eigenvalue matrix of the ID user channel,
interestingly it is observed that the proposed precoder achieves almost every
boundary points of the rate-energy region suggested in \cite{RZhang:13}.

{\it Notations:} Normal letters represent scalar quantities, boldface letters indicate vectors and boldface uppercase letters
designate matrices.
The superscripts $(\cdot)^T$, $(\cdot)^{H}$, and $\mathcal{E}[\cdot]$ stand for the transpose, the conjugate transpose,
and the expectation operators, respectively.
$\Iv_N$ is an $N \times N$ identity matrix.
In addition, $\textrm{Tr}\left(\Av\right)$, $|\Av|$, and $\Vert\Av\Vert_2$
indicate the trace, the determinant, and the matrix 2-norm of a matrix $\Av$, respectively.

\section{System Model \& Problem Formulation}\label{sec:System Model}

\begin{figure}
\begin{center}
\includegraphics[width=3.4in]{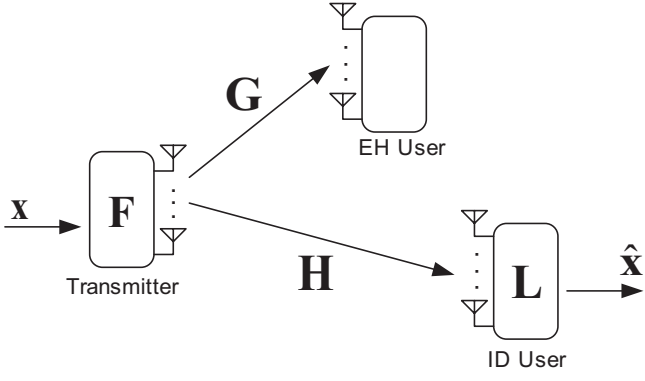}
\end{center}
\caption{A MIMO broadcast system for joint wireless information and energy transfer with linear transceiver \label{figure:SystemModel.eps}}
\end{figure}

As shown in Figure \ref{figure:SystemModel.eps}, we consider a wireless multi-antenna broadcast system
where one transmitter equipped with $N_T$ antennas transmits radio signals
to one EH user with $N_{\text{EH}}$ antennas and one ID user with $N_{\text{ID}}$ antennas at the same time and frequency.
Each user is chosen from the ``near-far'' based scheduling at the transmitter, i.e.,
the closest user to the transmitter is scheduled as a EH user and the other one, further away, as an ID user.
We assume that the two users are physically separated, but it can be easily extended to the co-located cases by applying a power splitting method in \cite{RZhang:13}.
In this paper, we assume quasi-static flat fading channels so that
the base-band channels from the transmitter to the EH and ID receivers are simply represented by complex matrices
$\Hv\in\mathbb{C}^{N_{\text{ID}}\times N_T}$ and $\Gv\in\mathbb{C}^{N_{\text{EH}}\times N_T}$, respectively.
It is also assumed that the transmitter knows all channel state information (CSI) $\Hv$ and $\Gv$, and each user knows the corresponding CSI.
Considering the spatial multiplexing scheme,
we transmit $N_S$ data-streams simultaneously using the input signal vector
$\xv\in\mathbb{C}^{N_S\times1}$ with $\mathcal{E}[\xv\xv^H]=\Iv_{N_S}$.

With these assumptions, the estimated signal $\hat{\xv}\in\mathbb{C}^{N_S\times1}$ at the ID user can be modeled by
\bea\label{eq:system model}
\hat{\xv}=\Lv(\Hv\Fv\xv+\nv),
\eea
where $\nv\in\mathbb{C}^{N_\text{ID}\times1}$ indicates the receiver noise with $\nv\sim\mathcal{CN}(\mathbf{0},\Iv_{N_{\text{ID}}})$,
$\Lv\in\mathbb{C}^{N_S\times N_{\text{ID}}}$ represents the linear receiver at the ID user, and
$\Fv\in\mathbb{C}^{N_T\times N_S}$ is the linear precoder at the transmitter,
which is subject to the power constraint $\text{Tr}(\Fv\Fv^H)\leq P_T$.
Then, the estimation error at the ID user can be defined as $\ev=\gamma^{-1}\hat{\xv}-\xv$
and the resulting weighted sum-MSE is given by $\text{MSE}=\text{Tr}(\Wv \mathcal{E}[\ev\ev^H])$. Here, the scaling parameter $\gamma^{-1}$ plays
a role of automatic gain control \cite{JJingon:07} which helps simplifying the derivation and $\Wv$ stands for a real diagonal weight matrix.
Note that without consideration of the EH user,
the MMSE optimum transmitter design which we denote by $\Fv_\text{ID}$
is well investigated in literature \cite{Sampath:01}.

In the meantime, the total harvested energy at the EH receiver, defined by $Q$, can be expressed as
\bea\label{eq:harvested energy}
Q=\delta \mathcal{E}[\Vert\Gv\Fv\xv\Vert^2]=\text{Tr}(\Fv^H\Gv^H\Gv\Fv)
\eea
where $\delta$ is a constant accounting for the harvesting efficiency and we assume $\delta=1$ for simplicity unless stated otherwise.
Let $\sqrt{g_1}$ and $\vv_{g,1}\in\mathbb{C}^{N_T\times1}$ denote the largest singular value of $\Gv$ and the corresponding right singular vector, respectively.
Then, it is known that the energy beamformer $\Fv_\text{EH}=\sqrt{P_T}[\vv_{g,1} ~\mathbf{0}_{N_T\times(N_S-1)}]$ achieves
the maximum energy $E_{\max}=P_T g_1$ \cite{JPark:13} \cite{RZhang:13}.
and we denote the achievable sum-MSE with this beamformer by $M_{EH}$.
Meanwhile, the sum-MSE and the energy attainable by the MMSE precoder $\Fv_\text{ID}$ are denoted by $M_{\min}$ and $E_\text{ID}$, respectively.

Now, consider the case where both the EH and ID users co-exist. Then, the achievable MSE-energy region is
\bea
\mathcal{R}_{M,E}\triangleq\Big\{(M,E):M\geq \text{Tr}(\Wv \mathcal{E}[\ev\ev^H]),~~~~~~~~~~~~~~~~~~~\nonumber\\
E\leq\text{Tr}(\Fv^H\Gv^H\Gv\Fv),~\text{Tr}(\Fv\Fv^H)\leq P_T\Big\}\nonumber,
\eea
and each boundary point of $\mathcal{R}_{M,E}$ can be obtained by solving the following SWIPT-WMMSE optimization problem
\bea\label{eq:optimization problem}
\min_{\gamma,\mathbf{F},\mathbf{L}}&&\text{Tr}(\Wv \mathcal{E}[\ev\ev^H])\nonumber\\
s.t. &&\text{Tr}(\Fv^H\Gv^H\Gv\Fv)\geq\bar{E},~\text{Tr}(\Fv\Fv^H)\leq P_T,
\eea
where $E_\text{ID}\leq\bar{E}\leq E_{\max}$.
This problem is convex over the transmitter (receiver) when the receiver (transmitter) and $\gamma$ are given, and
strictly quasi-convex with respect to $\gamma$ when both $\Fv$ and $\Lv$ are given.
However, it is generally non-convex on $\{\gamma,\Fv,\Lv\}$.
Our goal of the paper is to find efficient solutions to problem (\ref{eq:optimization problem}) and identify the best possible MSE-energy region $\mathcal{R}_{M,E}$.

\section{Weighted MMSE Transceiver Designs}\label{sec:Optimal Transceiver Design}
In this section, we solve the problem (\ref{eq:optimization problem}) based on the Karush-Kuhn Tucker (KKT) conditions.
First, we propose the optimum precoder structure in terms of the WMMSE,
and then suggest alternative optimization process between the transmitter and the ID receiver.
Several interesting observations will also be made when the ID user is equipped with a single antenna.

Using the expression of $\ev$ and the assumptions made in the previous section,
we formulate the Lagrangian as
\bea\label{eq:Lagrangian}
\mathcal{L}(\lambda, \mu, \gamma, \Fv, \Lv)~~~~~~~~~~~~~~~~~~~~~~~~~~~~~~~~~~~~~~~~~~~~~~~\nonumber\\
=\text{Tr}\big(\Wv(\gamma^{-1}\Lv\Hv\Fv-\Iv_{N_S})(\gamma^{-1}\Lv\Hv\Fv-\Iv_{N_S})^H\big)~~~~~\nonumber\\
+\text{Tr}\big(\gamma^{-2}\Wv\Lv\Lv^H\big)-\lambda\big(\text{Tr}(\Fv^H\Gv^H\Gv\Fv)-\bar{E}\big)\nonumber\\
+\mu\big(\text{Tr}(\Fv^H\Fv)-P_T\big).~~~~~~~~~~~~~~~~~~~~~~~~~~~~
\eea
Then, the following KKT conditions provide necessary conditions for optimality:
\bea
\label{eq:KKT1}
\Lv\Hv\Fv\Fv^H\Hv^H+\Lv=\gamma\Fv^H\Hv^H\\
\label{eq:KKT2}
\Hv^H\Lv^H\Wv\Lv\Hv\Fv-\bar{\lambda}\Gv^H\Gv\Fv+\bar{\mu}\Fv=\gamma\Hv^H\Lv^H\Wv\\
\label{eq:KKT3}
\text{Tr}\left(\Wv\Lv\Hv\Fv\Fv^H\Hv^H\Lv^H+\Wv\Lv\Lv^H\right)~~~~~\nonumber\\
=\gamma\text{Tr}\left(\Fv^H\Hv^H\Lv^H\Wv\right)\\
\label{eq:KKT4}
\lambda\geq0;\mu\geq0;\text{Tr}(\Fv^H\Gv^H\Gv\Fv)\geq\bar{E};\text{Tr}(\Fv^H\Fv)\leq P_T\\
\label{eq:SlackT}
\bar{\mu}\left(\text{Tr}(\Fv^H\Fv)-P_T\right)=0\\
\label{eq:SlackE}
\bar{\lambda}\left(\text{Tr}(\Fv^H\Gv^H\Gv\Fv)-\bar{E}\right)=0,
\eea
where $\bar{\lambda}\triangleq\lambda\gamma^2$, $\bar{\mu}\triangleq\mu\gamma^2$, and
equations (\ref{eq:KKT1})-(\ref{eq:KKT3}) are derived from the zero gradient conditions by using some rules of differentiation \cite{AHjorungnes:07}
and (\ref{eq:SlackT}) and (\ref{eq:SlackE}) come from the complement slackness conditions.

Now, let us define two matrices that will be used for our derivations as
\bea\label{eq:definitionZ}
\Yv&=&\bigg(\mathbf{H}^H\mathbf{L}^H\mathbf{W}\mathbf{L}\mathbf{H}+\frac{\text{Tr}(\mathbf{W}\mathbf{L}\mathbf{L}^H)}{P_T}\Iv_{N_T}\bigg)\nonumber\\
\Zv&=&\bigg(\Gv^H\Gv-\frac{\bar{E}}{P_T}\Iv_{N_T}\bigg).\nonumber
\eea
Here, $\Yv$ is a positive definite matrix, but $\Zv$ is in general positive indefinite which makes the analysis challenging.
Based on these results, each of the transmitter and the receiver can be optimized separately as in the following theorems.

\begin{Theorem}\label{Theorem:Theorem1}
{\it When $\Lv$ is given arbitrarily, the solutions of problem (\ref{eq:optimization problem}) are attained by
\bea\label{eq:theorem1}
\hat{\Fv}&=&\hat{\gamma}\bar{\Fv}(\bar{\lambda})=\hat{\gamma}\left(\Yv-\bar{\lambda}\Zv\right)^{-1}\Hv^H\Lv^H\Wv,
\eea
where $\hat{\gamma}=\sqrt{\frac{P_T}{\text{Tr}(\bar{\Fv}(\bar{\lambda})\bar{\Fv}(\bar{\lambda})^H)}}$.
Defining $J(x)=\text{Tr}\left(\bar{\Fv}(x)^H\Zv\bar{\Fv}(x)\right)$,
if $J(0)\geq0$, we have $\bar{\lambda}=0$ and
otherwise, a unique optimal $\bar{\lambda}$ satisfying $J(\bar{\lambda})=0$
can always be found over the range $0<\bar{\lambda}<1/\kappa$ by simple line search methods where
$\kappa\triangleq\Vert\Zv\Yv^{-1}\Vert_2^2$.}
\end{Theorem}
\begin{IEEEproof}
Let us first consider the Lagrange dual function $g(\lambda,\mu)=\min_{\gamma,\mathbf{F}}\mathcal{L}(\lambda, \mu, \gamma, \Fv, \Lv)$
with fixed $\lambda$, $\mu$, and $\Lv$.
Then, ignoring constant terms, the problem of minimizing $\mathcal{L}$ over $\Fv$ and $\gamma$ is equivalently
\bea\label{eq:Lagrangian2}
\min_{\gamma,\mathbf{F}}~\gamma^{-2}\big(\text{Tr}(\Fv^H\Kv\Fv)
-2\gamma\text{Tr}(\Re(\Wv\Lv\Hv\Fv))\big),
\eea
where $\Kv\triangleq\Hv^H\Lv^H\Wv\Lv\Hv-\bar{\lambda}\Gv^H\Gv+\bar{\mu}\Iv_{N_T}$.
Now, suppose that at least one eigenvalue of $\Kv$ is a non-positive real number
with corresponding eigenvector $\vv\in\mathbb{C}^{N_T\times1}$. Then, it is easy to show
that problem (\ref{eq:Lagrangian2}) becomes unbounded below with $\Fv=[\vv ~\mathbf{0}_{N_T\times(N_S-1)}]$ as $\gamma\rightarrow0^+$.
Therefore, to obtain a bounded optimal value of problem (\ref{eq:optimization problem}),
the optimal $\bar{\lambda}$ and $\bar{\mu}$ must always be chosen such that $\Kv\succ0$.

Assuming $\Kv\succ0$, we now obtain from KKT condition (\ref{eq:KKT2}) that $\hat{\Fv}=\gamma\Kv^{-1}\Hv^H\Lv^H\Wv$.
Also, from (\ref{eq:KKT2})-(\ref{eq:KKT4}), it follows
\bea
\gamma\text{Tr}(\Fv^H\Hv^H\Lv^H\Wv)~~~~~~~~~~~~~~~~~~~~~~~~~~~~~~~~~~~~~~~\nonumber\\
=\text{Tr}(\Fv^H\Hv^H\Lv^H\Wv\Lv\Hv\Fv-\bar{\lambda}\Fv^H\Gv^H\Gv\Fv+\bar{\mu}\Fv^H\Fv)\nonumber\\
=\text{Tr}(\Fv^H\Hv^H\Lv^H\Wv\Lv\Hv\Fv)-\bar{\lambda}\bar{E}+\bar{\mu}P_T~~~~~~~~~~~~~~~\nonumber\\
=\text{Tr}(\Wv\Lv\Hv\Fv\Fv^H\Hv^H\Lv^H)+\text{Tr}(\Wv\Lv\Lv^H),~~~~~~~~~~\nonumber
\eea
where the second equality is due to (\ref{eq:KKT4}) and the last equality follows from (\ref{eq:KKT3}).
Comparing the last two, we can show that $\bar{\mu}=\frac{\bar{\lambda}\bar{E}+\text{Tr}(\mathbf{W}\mathbf{L}\mathbf{L}^H)}{P_T}$,
which allows us to reduce $J(\bar{\lambda},\bar{\mu})$ to a simple one-dimensional dual function $J(\bar{\lambda})$.
Since $\bar{\mu}>0$, we also have $\text{Tr}(\Fv^H\Fv)=P_T$, i.e., $\hat{\gamma}=\left(P_T/\text{Tr}(\bar{\Fv}(\bar{\lambda})\bar{\Fv}(\bar{\lambda})^H)\right)^{1/2}$ from (\ref{eq:SlackT}).

Before we further proceed with our proof, let us show the following two useful lemmas.
The proofs can be found in Appendix.
\begin{Lemma}\label{Lemma:Lemma1}
{\it The optimal $\bar{\lambda}$ takes a value over the range $0\leq\bar{\lambda}<1/\kappa$.}
\end{Lemma}
\begin{Lemma}\label{Lemma:Lemma2}
{\it $J(x)$ is a monotonic increasing function for $0\leq x<1/\kappa$ and $\lim_{x\rightarrow1/\kappa}J(x)>0$.}
\end{Lemma}

With the aid of Lemma \ref{Lemma:Lemma1} and \ref{Lemma:Lemma2}, the optimal dual variable $\bar{\lambda}$ which we denote by $\hat{\lambda}$ can be obtained as follows:
We first see from the KKT condition (\ref{eq:SlackE}) that
$\hat{\lambda}\left(\text{Tr}(\gamma^2\bar{\Fv}(\hat{\lambda})^H\Gv^H\Gv\bar{\Fv}(\hat{\lambda}))-\bar{E}\right)=0$
must be satisfied, which implies that
$\hat{\lambda}J(\hat{\lambda})=0$.
Therefore, if $J(0)\geq0$, we attain $\hat{\lambda}=0$, because $J(\bar{\lambda})>0$ for $\bar{\lambda}>0$.
In contrast, if $J(0)<0$, one can find a unique $\hat{\lambda}\neq0$ satisfying $J(\hat{\lambda})=0$ by
simple line search methods over $0<\bar{\lambda}<1/\kappa$, since $J(1/\kappa)>0$.
It is worthwhile noting that $J(\hat{\lambda})$ must always be $0$ or positive, because otherwise,
the target energy $\bar{E}$ is not attainable at the EH receiver.
Finally, as the solution satisfying the
KKT conditions (\ref{eq:KKT2})-(\ref{eq:SlackE}) is unique, the resulting precoder (\ref{eq:theorem1}) is also sufficient
for the optimality, and the proof is completed.
\end{IEEEproof}

\begin{Theorem}\label{Theorem:Theorem2}
{\it When $\gamma$ and $\Fv$ are given, the solution of problem (\ref{eq:optimization problem}) is expressed as}
\bea\label{eq:theorem2}
\hat{\Lv}=\gamma\Fv^H\Hv^H(\Hv\Fv\Fv^H\Hv^H+\Iv_{N_\text{ID}})^{-1}
\eea
\end{Theorem}
\begin{IEEEproof}
The proof is trivial; thus omitted.
\end{IEEEproof}

\begin{algorithm}
\caption{ WMMSE Optimization Method}
\begin{algorithmic}
\FOR{$j=1:N_G$}
    \STATE Initialize  $~\bar{\Fv}_j$ and $\gamma_j=\big(P_T/\text{Tr}(\bar{\Fv}_j\bar{\Fv}_j^H)\big)^{1/2}$
    \REPEAT
        \STATE Compute $\Lv_j$ using (\ref{eq:theorem2}) with given $\bar{\Fv}_j$ and $\gamma_j$
        \STATE Update $\Fv_j=\gamma_j\bar{\Fv}_j$ using (\ref{eq:theorem1}) with given $\Lv_j$
    \UNTIL{ $\Fv_j$ converge to the prescribed accuracy}
\ENDFOR
\STATE Select the best solution among $\big\{(\Fv_j, \Lv_j)|j=1,\ldots,N_G\big\}$, which shows the minimum sum-MSE
\end{algorithmic}
\label{algorithm:algorithm1}
\end{algorithm}

Based on results in Theorem \ref{Theorem:Theorem1} and \ref{Theorem:Theorem2}, the problem (\ref{eq:optimization problem})
can be solved in an alternating fashion between the transmitter and the receiver as summarized in Algorithm \ref{algorithm:algorithm1}.
The sum-MSE monotonically decreases by each transmit and receive processing update.
Therefore, the inner-iteration of Algorithm \ref{algorithm:algorithm1} guarantees the convergence at least to a local minimum.
However, due to joint non-convexity of problem (\ref{eq:optimization problem}),
numerous initial points which we denote by $N_G$ are employed so that
the resulting local minimum gets closer to the global minimum.
This requires additional outer loop iterations.

Next, let us examine the optimal beamforming solutions with $N_{\text{ID}}=N_S=1$ for the special cases where
the ID receiver or both the ID and EH receivers have a single antenna.
In this case, the transmit matrix $\Fv$ reduces to a column vector $\fv\in\mathbb{C}^{N_T\times1}$,
the receive matrix $\Lv$ is equivalent to a scalar value $l\in\mathbb{C}$, and
the weight matrix $\Wv$ can be ignored without loss of generality.

\begin{Corollary}\label{Corollary:Corollary1}
{\it In the case of MISO channel from the transmitter to the ID receiver, i.e.,
$\Hv\equiv\hv^H$ with $\hv\in\mathbb{C}^{N_T\times1}$,
the MMSE optimal beamformer, which is independent of the receiver $l$, is given by
\bea\label{eq:corollary1}
\hat{\fv}=\hat{\gamma}\bar{\fv}=\hat{\gamma}\Av^{-1}\hv\Big(\frac{1}{1+\hv^H\Av^{-1}\hv}\Big),
\eea
where $\hat{\gamma}=\sqrt{P_T/\Vert\bar{\fv}\Vert^2}$ and
$\Av\triangleq P_T^{-1}\Iv_{N_T}-\bar{\lambda}\Zv$ where $\bar{\lambda}$ is obtained as follows:
If $J(0)\geq0$, we set $\bar{\lambda}=0$ and otherwise, we find a unique $\bar{\lambda}$ satisfying $J(\bar{\lambda})=0$ by simple line search methods
over $0<\bar{\lambda}<1/\zeta$ with $\zeta=\Vert\Zv(\mathbf{h}\mathbf{h}^H+\frac{1}{P_T}\Iv_{N_T})^{-1}\Vert_2^2$.}
\end{Corollary}
\begin{IEEEproof}
Assuming $N_S=N_\text{ID}=1$ and following similar approaches as in Theorem \ref{Theorem:Theorem1},
the KKT conditions (\ref{eq:KKT1})-(\ref{eq:SlackE}) lead to the solution
\bea\label{eq:beamformer1}
\hat{\fv}=\gamma\Big(\hv\hv^H+\frac{1}{P_T}\Iv_{N_T}-\nu\Zv\Big)^{-1}\hv l^{-1},
\eea
where $\nu\triangleq|l|^{-2}\bar{\lambda}$ and $\gamma=\sqrt{P_T/\Vert\bar{\fv}\Vert^2}$
and $\nu$ is chosen to satisfy $\nu J(\nu)=0$ over the range $0\leq\nu<1/\zeta$.
Since a particular choice of $l$ has no influence on the optimal value of $\nu$ and
the resulting MSE of (\ref{eq:beamformer1}), i.e., $E=(1+\mathbf{h}^H\mathbf{A}^{-1}\mathbf{h})^{-1}$ \cite{Joham:05} is independent of $l$,
setting $l=1$ will not hurt the generality.
Finally, by invoking some matrix inversion lemma, (\ref{eq:beamformer1}) is alternatively expressed as
$\fv=\gamma\Av^{-1}\hv(1+\hv^H\Av^{-1}\hv)^{-1}$,
and the proof is completed.
\end{IEEEproof}

\begin{table*}[!htp]
\centering \caption{System Parameters}
\begin{tabular}{|c||c|}
\hline
Noise Power Spectral Density & $-100$ dBm/Hz\\
\hline
Energy Conversion Efficiency & $50\%$ ($\delta=0.5$)\\
\hline
Pathloss Exponent & $3$ \\
\hline
Signal Bandwidth & $10$ MHz \\
\hline
Distance: Tx - ID receiver & $d_h$(m)\\
\hline
Distance: Tx - EH receiver & $d_g$(m)\\
\hline
$\begin{array}{c}\text{Channel Realization}\vspace{-0.1cm}\\\text{(MATLAB Code)}\end{array}$ &
$\begin{array}{c}\text{randn(`state',$\theta$)};~~~~~~~~~~~~~~~~~~~~~~~~~~~~~~~~~~~~~~~~~~~~~~~~~\\
\widetilde{\Hv}=(\text{randn}(N_{\text{ID}},N_T) + j*\text{randn}(N_{\text{ID}},N_T))/\text{sqrt}(2);~\\
\widetilde{\Gv}=(\text{randn}(N_{\text{EH}},N_T) + j*\text{randn}(N_{\text{EH}},N_T))/\text{sqrt}(2);\end{array}$\\
\hline
\end{tabular}
\label{table:table1}
\end{table*}

In fact, when we deal with a single stream transmission, i.e., $N_S=1$,
the MMSE beamforming also maximizes the end-to-end SNR which leads to the maximum information rate \cite{Palomar:03} \cite{Joham:05}.
Therefore, our beamforming strategy in Corollary \ref{Corollary:Corollary1} is essentially equivalent
to one in \cite[Corollary 3.1]{RZhang:13} which was developed for the rate maximization.
Interestingly, it is shown that while the previous approach \cite{RZhang:13} requires a complicated optimization tool such as the ellipsoid method to obtain the solution,
our method enables us to achieve the maximum rate by a semi-closed-form solution that can be solved by a simple bisection method;
thus, providing more insights into the system as well as complexity reduction.

\section{Numerical Results}

In this section, we provide several interesting observations
by comparing the numerical performance of the two MIMO BC SWIPT design criteria: WMMSE and the information rate \cite{RZhang:13}.
Key system parameters are summarized in Table \ref{table:table1}.
Throughout our simulation, the channels are generated from Rayleigh fading pathloss model,
$\Hv=d_h^{-3/2}\widetilde{\Hv}$ and $\Gv=d_g^{-3/2}\widetilde{\Gv}$
where each element of $\widetilde{\Hv}$ and $\widetilde{\Gv}$
is drawn from the i.i.d. CSCG normal distribution according to the MATLAB code with the seed number $\theta$ as in Table \ref{table:table1}.
Also, we set $N_T=N_S=N_\text{ID}=N_\text{EH}=4$ and $d_h=d_g=10$(m) for the ease of the presentation.
Let us assume that $P_T=20$ dBm ($100mW$).
Then, since the noise power is $-100$ dBm/Hz $\times$ $100$MHz=$1\mu W$,
the per-antenna SNR at the ID user becomes $20-30-(-30) =20$ dB, which
is equivalent to the standard signal model with $P_T=10^5$ in (\ref{eq:system model}).
Thus, $1$ energy unit at the EH user in (\ref{eq:harvested energy}) amounts to $1\mu W$ in our simulation setting.

\begin{figure}
\begin{center}
\includegraphics[width=3.5in]{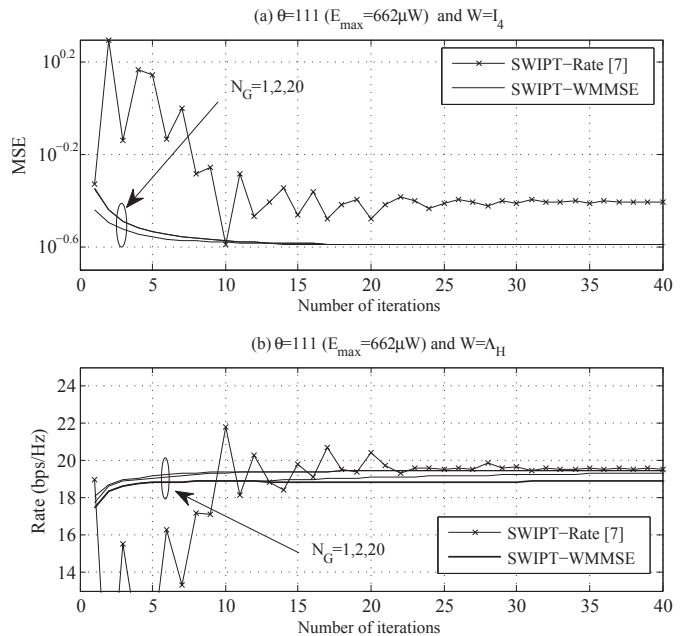}
\end{center}
\caption{Sum-MSE and rate performance comparison of the two MIMO BC SWIPT methods with $\bar{E}=E_{\max}/2$}
\label{figure:ConvergenceRate1.eps}
\end{figure}
\begin{figure}
\begin{center}
\includegraphics[width=3.4in]{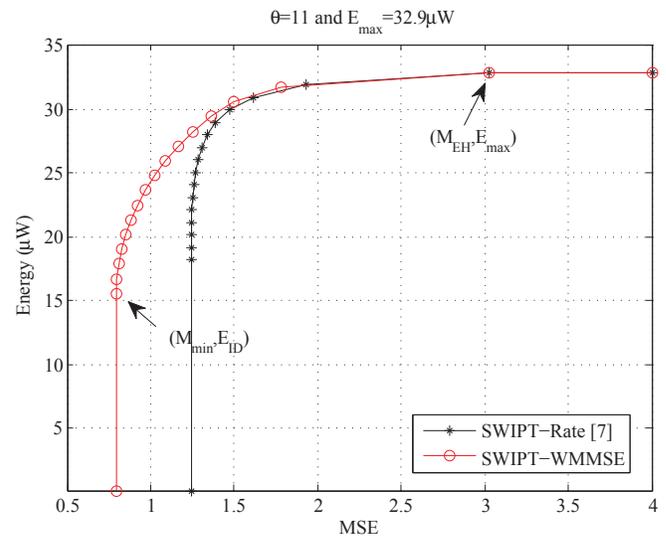}
\end{center}
\caption{Achievable MSE-energy region of the two MIMO BC SWIPT methods with $P_T=10$mW and $\Wv=\Iv_{4}$}
\label{figure:MSE-energy.eps}
\end{figure}

Figure \ref{figure:ConvergenceRate1.eps}-(a) and 2-(b) illustrate the convergence trends of the proposed scheme with $P_T=100$mW
in terms of both the sum-MSE and the information rate with weight matrices $\Wv=\Iv_4$ and $\Wv=\mathbf{\Lambda}_H$, respectively,
where $\mathbf{\Lambda}_H$ indicates the eigenvalue matrix of $\Hv^H\Hv$.
In each iteration, the sum-MSE ($M$) and the rate ($R$) are computed from
$M=\text{Tr}(\Wv(\Fv^H\Hv^H\Hv\Fv+\Iv_{N_T})^{-1})$ and $R=\log|\Fv^H\Hv^H\Hv\Fv+\Iv_{N_T}|$
under the assumption that the ID user employs the MMSE receiver (\ref{eq:theorem2}) and the maximum likelihood receiver, respectively.
Due to the non-convexity of problem (\ref{eq:optimization problem}), the resulting performance
depends on the number of initial points and iterations, especially when the weight matrix is non-identity.
Nevertheless, it is seen that $20$ random initial points and $10$ number of iterations seem to be
sufficient to achieve the converged performance for our scheme.
It is somewhat natural that the proposed MMSE scheme obtains improved MSE performance over the {\it ``SWIPT-Rate''} scheme \cite{RZhang:13}.
However, it is particularly interesting to observe that our solution achieves the optimum rate as well when the weight factors are properly chosen,
which implies that the rate maximizing precoder design can be regarded as a special case of our WMMSE design\footnote{Similar observation has been made in
point-to-point MIMO systems without energy transfer \cite{Sampath:01}}.
Extensive computer simulations confirm that $\Wv=\mathbf{\Lambda}_H$ is the best choice for maximizing the rate, but
rigorous proof remains open.

\begin{figure}
\begin{center}
\includegraphics[width=3.4in]{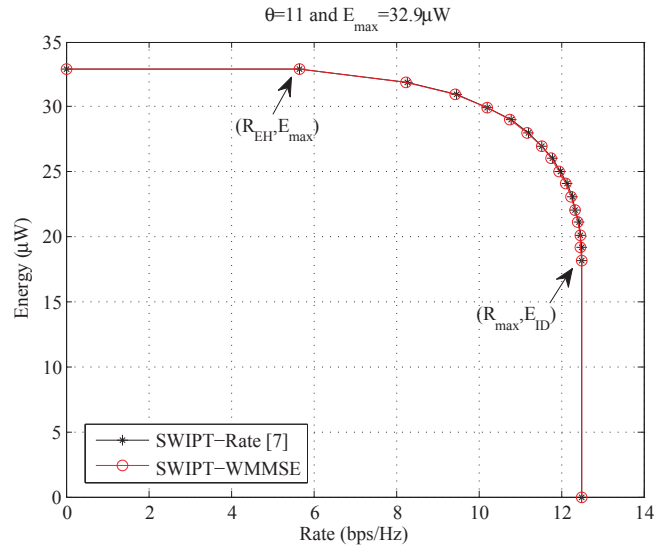}
\end{center}
\caption{Achievable Rate-energy region of the two MIMO BC SWIPT methods with $P_T=10$mW and $\Wv=\mathbf{\Lambda}_H$}
\label{figure:Rate-energy.eps}
\end{figure}

Figures \ref{figure:MSE-energy.eps} and \ref{figure:Rate-energy.eps} illustrate the boundary points of
the achievable MSE-energy and rate-energy regions, respectively.
20 random initial points are used for the proposed SWIPT-WMMSE scheme.
It is shown that our solution achieves the outstanding MSE-energy region compared to the SWIPT-Rate scheme \cite{RZhang:13},
which leads to the BER performance improvement as is shown later.
In addition, Figure \ref{figure:Rate-energy.eps} shows that when $\Wv=\mathbf{\Lambda}_H$,
the WMMSE precoder achieves almost every boundary points of the optimal rate-energy region;
thus, confirming the previous observation made in Figure \ref{figure:ConvergenceRate1.eps}.
We would like to note that although here we presented simulations results for a sample channel ($\theta=11$),
the trend does not change for all channel realizations.

\begin{figure}
\begin{center}
\includegraphics[width=3.4in]{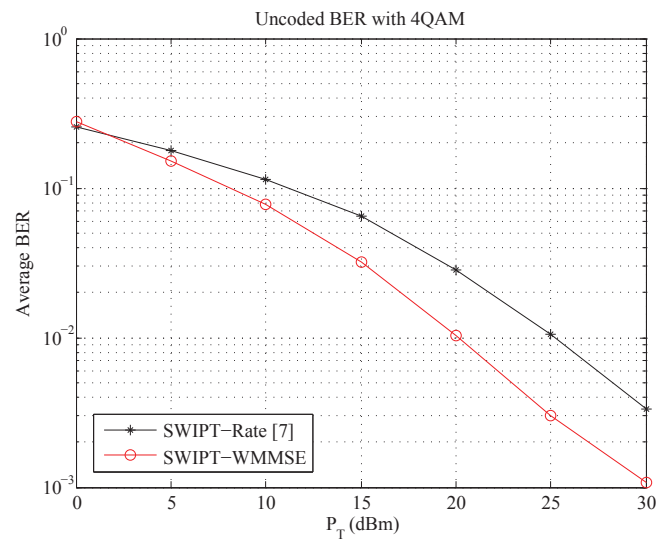}
\end{center}
\caption{BER performance comparison of the two MIMO BC SWIPT methods with $\Wv=\Iv_4$ and $\bar{E}=E_{\max}/2$}
\label{figure:BER.eps}
\end{figure}

Figure \ref{figure:BER.eps} compares the uncoded BER performance of the two MIMO BC SWIPT systems with 4QAM.
We averaged over $10^5$ random channel realizations and set $\bar{E}=E_{\max}/2$ in each realization so that the EH user
harvests at least $50$\% of the maximum energy allowed for a given channel.
The figure shows that the proposed MMSE scheme with $\Wv=\Iv_4$ obtains approximately $5$ dB gain over the SWIPT-Rate scheme.
This is because the rate-based precoder allocates more resource to the stronger sub-channel,
while the MMSE precoder neutralizes the channel gains across sub-channels.
On the contrary, for the same reason, the rate-based design may exhibit better error performance
when the transmitter adopts sufficiently strong channel codes.
The proposed WMMSE precoder will softly bridge two extreme cases by adjusting the weight matrix $\Wv$.

\section{Conclusion}

In this paper, we investigated efficient linear transceiver designs for MIMO BC SWIPT systems
which minimizes the weighted sum-MSE of the ID-user while satisfying the energy harvesting constraint of the EH-user.
First, we identified the optimal precoder structure as a closed-form solution that can be obtained by simple bisection methods.
Then, it was shown that the derived precoder achieves the maximum SNR at the ID user with reduced complexity when the ID user has a single antenna.
Second, we suggested alternative updating process to obtain the joint WMMSE solution with the ID receiver.
Although the problem is non-convex, it is observed from simulation results that
our solution with multiple initial points approaches the best possible MSE-energy tradeoff region.
It was also interesting to observe that the proposed WMMSE precoder with $\Wv=\mathbf{\Lambda}_{H}$ achieves
almost every boundary points of the optimal rate-energy region.

\section{Acknowledgement}
This work was supported in part by FP7 project PHYLAWS (EU FP7-ICT 317562).

\appendix

\subsection{Proof of Lemma \ref{Lemma:Lemma1}} \label{Appendix:Appendix 1}
Using $\bar{\mu}=\frac{\bar{\lambda}\bar{E}+\text{Tr}(\mathbf{W}\mathbf{L}\mathbf{L}^H)}{P_T}$,
the feasibility condition $\Kv\succ0$ is equivalently
$\Yv\succ\bar{\lambda}\Zv$,
which holds if and only if $\bar{\lambda}<1/\kappa$ where $\kappa$ indicates the maximum eigenvalue of a matrix $\Zv\Yv^{-1}$.
Note that the maximum eigenvalue of a matrix is equivalent to its matrix two norm \cite{Golub:96}.
Therefore, combining with the condition $\lambda\geq0$ in (\ref{eq:KKT4}), we obtain Lemma \ref{Lemma:Lemma1}.

\subsection{Proof of Lemma \ref{Lemma:Lemma2}} \label{Appendix:Appendix 2}
On the one hand, we first set the derivative of the function $J(x)$ with respect to $x$ as
\bea\label{eq:derivative}
\frac{\partial J(x)}{\partial x}&=&\text{Tr}\bigg(\Big(\frac{\partial J(x)}{\partial \bar{\Fv}(x)}\Big)^T\frac{\partial \bar{\Fv}(x)}{\partial x}\bigg)\nonumber\\
&=&\text{Tr}\bigg(\gamma^2[\Zv\bar{\Fv}(x)]^H\frac{\partial \bar{\Fv}(x)}{\partial x}\bigg).
\eea
In addition, taking a derivative at both sides of the equation (\ref{eq:KKT2}), i.e., $\Kv\bar{\Fv}(x)=\Hv^H\Lv^H\Wv$, we have
\bea
\frac{\partial}{\partial x}
\left[\Kv\bar{\Fv}(x)\right]
=-\Zv\bar{\Fv}(x)+\Kv\frac{\partial \bar{\Fv}(x)}{\partial x}=0,\nonumber
\eea
which gives us $\Zv\bar{\Fv}(x)=\Kv\frac{\partial \bar{\Fv}(x)}{\partial x}$.
Then, plugging this result back into (\ref{eq:derivative}), it follows
\bea
\frac{\partial J(x)}{\partial x}=\text{Tr}\left(\gamma^2\left(\frac{\partial \bar{\Fv}(x)}{\partial x}\right)^H\Kv\left(\frac{\partial \bar{\Fv}(x)}{\partial x}\right)\right),\nonumber
\eea
which implies that as long as $\Kv\succ0$ (or $x<1/\kappa$),
$J(x)$ is a monotonic increasing function of $x$.

On the other hand, applying the obtained precoder (\ref{eq:theorem1}) to $J(x)$, it is rewritten by
\bea
J(x)=\text{Tr}\big(\Wv^H\Lv^H\Hv^H(\Yv-x\Zv)^{-1}\Zv(\Yv-x\Zv)^{-1}\Hv\Lv\Wv\big)\nonumber\\
=\text{Tr}\big(\Wv^H\Lv^H\Hv^H(\Iv_{N_T}-x\Zv\Yv^{-1})^{-1}~~~~~~~~~~~~~~~~~~\nonumber\\
\times\Yv^{-1}\Zv(\Yv-x\Zv)^{-1}\Hv\Lv\Wv\big).~~~~~~~~~~~~~~\nonumber
\eea
This result implies that as $x\rightarrow 1/\kappa$, $J(x)$ will have an infinitely large number (+$\infty$ or $-\infty$)
due to the inverse operation of a rank deficient matrix $\Iv_{N_T}-\kappa^{-1}\Zv\Yv^{-1}$.
As shown previously, $J(x)$ is increasing function over $x<1/\kappa$, and thus
we have $\lim_{x\rightarrow 1/\kappa}J(x)\rightarrow+\infty$, and the proof is concluded.

\bibliographystyle{ieeetr}

\input{bibliography.filelist}



\end{document}